# Congestion Control and Traffic Management in ATM Networks: Recent Advances and A Survey

Raj Jain
Department of Computer and Information Science
The Ohio State University
Columbus, OH 43210
Email: Jain@ACM.Org
Draft Version: August 13, 1996

**Abstract**

Congestion control mechanisms for ATM networks as selected by the ATM Forum traffic management group are described. Reasons behind these selections are explained. In particular, selection criterion for selection between rate-based and credit-based approach and the key points of the debate between these two approaches are presented. The approach that was finally selected and several other schemes that were considered are described.

## 1 Introduction

Future high speed networks are expected to use the Asynchronous Transfer Mode (ATM) in which the information is transmitted using short fixed-size cells consisting of 48 bytes of payload and 5 bytes of header. The fixed size of the cells reduces the variance of delay making the networks suitable for integrated traffic consisting of voice, video, and data. Providing the desired quality of service for these various traffic types is much more complex than the data networks of today. Proper traffic management helps ensure efficient and fair operation of networks in spite of constantly varying demand. This is particularly important for the data traffic which has very little predictability and, therefore, cannot reserve resources in advance as in the case of voice telecommunications networks.

Traffic management is concerned with ensuring that users get their desired quality of service. The problem is especially difficult during periods of heavy load particularly if the traffic demands cannot be predicted in advance. This is why congestion control, although only a part of the traffic management issues, is the most essential aspect of traffic management.

Congestion control is critical in both ATM and non-ATM networks. When two bursts arrive simultaneously at a node, the queue lengths may become large very fast resulting in buffer overflow. Also, the range of link speeds is growing fast as higher speed links are being introduced in slower networks of the past. At the points, where the total input rate is larger than the output link capacity, congestion becomes a problem.

The protocols for ATM networks began being designed in 1984 when the Consultative Committee on International Telecommunication and Telegraph (CCITT) - a United Nations Organization responsible for telecommunications standards - selected ATM as the paradigm for its broadband integrated service digital networks (B-ISDN). Like most other telecommunications standards, the ATM standards specify the interface between various networking components. A principal focus of these standards is the user-network interface (UNI), which specifies how a computer system (which is owned by a user) should communicate with a switch (which is owned by the network service provider).

ATM networks are connection-oriented in the sense that before two systems on the network can communicate, they should inform all intermediate switches about their service requirements and traffic parameters. This is similar to the telephone networks where a circuit is setup from the calling party to the called party. In ATM networks, such circuits are called virtual circuits (VCs). The connections allow the network to guarantee the quality of service by limiting the number of VCs. Typically, a user declares key service requirements at the time of connection set up, declares the traffic parameters and may agree to control these parameters dynamically as demanded by the network.

This paper is organized as follows. Section 2 defines various quality of service attributes. These attributes help define various classes of service in Section 3. We then provide a general overview of congestion control mechanisms in Section 4 and describe the generalized cell algorithm in Section 4.1. Section 5 describes the criteria that were set up for selecting the final approach. A number of congestion schemes are described briefly in Section 6 with the credit-based and rate-based approaches described in more details in Sections 7 and 8. The debate that lead to the eventual selection of the rate-based approach is presented in Section 9.

The description presented here is not intended to be a precise description of the standard. In order to make the concept easy to understand, we have at times simplified the description. Those interested in precise details should consult the standards documents, which are still being developed as of this writing in January 1995.

## 2 Quality of Service (QoS) and Traffic Attributes

While setting up a connection on ATM networks, users can specify the following parameters related to the input traffic characteristics and the desired quality of service.

1. **Peak Cell Rate (PCR)**: The maximum instantaneous rate at which the user will transmit.

2. **Sustained Cell Rate (SCR)**: This is the average rate as measured over a long interval.

3. **Cell Loss Ratio (CLR)**: The percentage of cells that are lost in the network due to error and congestion and are not delivered to the destination.

$$\text{Cell Loss Ratio} = \frac{\text{Lost Cells}}{\text{Transmitted Cells}}$$

Each ATM cell has a "Cell Loss Priority (CLP)" bit in the header. During congestion, the network first drops cells that have CLP bit set. Since the loss of CLP=0 cell is more harmful to the operation of the application, CLR can be specified separately for cells with CLP=1 and for those with CLP=0.

4. **Cell Transfer Delay (CTD)**: The delay experienced by a cell between network entry and exit points is called the cell transfer delay. It includes propagation delays, queueing delays at various intermediate switches, and service times at queueing points.

5. **Cell Delay Variation (CDV)**: This is a measure of variance of CTD. High variation implies larger buffering for delay sensitive traffic such as voice and video. There are multiple ways to measure CDV. One measure called "peak-to-peak" CDV consists of computing the difference between the $(1-\alpha)$-percentile and the minimum of the cell transfer delay for some small value of $\alpha$.

6. **Cell Delay Variation Tolerance (CDVT) and Burst Tolerance (BT)**: For sources transmitting at any given rate, a slight variation in the inter-cell time is allowed. For example, a source with a PCR of 10,000 cells per second should nominally transmits cells every 100 $\mu$s. A leaky bucket type algorithm called "Generalized Cell Rate Algorithm (GCRA)" is used to determine if the variation in the inter-cell times is acceptable. This algorithm has two parameters. The first parameter is the nominal inter-cell time (inverse of the rate) and the second parameter is the allowed variation in the inter-cell time. Thus, a GCRA(100$\mu$s, 10$\mu$s), will allow cells to arrive no more than 10 $\mu$s earlier than their nominal scheduled time. The second parameter of the GCRA used to enforce PCR is called Cell Delay Variation Tolerance (CDVT) and of that used to enforce SCR is called Burst Tolerance (BT).

7. **Maximum Burst Size (MBS)**: The maximum number of back-to-back cells that can be sent at the peak cell rate but without violating the sustained cell rate is called maximum burst size (MBS). It is related to the PCR, SCR, and BT as follows:

$$\text{Burst Tolerance} = (\text{MBS} - 1)\left(\frac{1}{\text{SCR}} - \frac{1}{\text{PCR}}\right)$$

Since MBS is more intuitive than BT, signalling messages use MBS. This means that during connection setup, a source is required to specify MBS. BT can be easily calculated from MBR, SCR, and PCR.

Note that PCR, SCR, CDVT, BT, and MBS are input traffic characteristics and are enforced by the network at the network entry. CLR, CTD, and CDV are qualities of service provided by the network and are measured at the network exit point.

8. **Minimum Cell Rate (MCR)**: The is the minimum rate desired by a user.

Only the first six of the above parameters were specified in UNI version 3.0. MCR has been added recently and will appear in the next version of the traffic management document.

Table 1: ATM Layer Service Categories

| Attribute | CBR | rt-VBR | nrt-VBR | UBR | ABR |
|---|---|---|---|---|---|
| PCR and CDVT[5] | Specified | | | Specified[3] | Specified[4] |
| SCR, MBS, CDVT[5,6] | n/a | Specified | | n/a | |
| MCR[5] | n/a | | | n/a | Specified |
| Peak-to-peak CDV | Specified | Specified | Unspecified | Unspecified | Unspecified |
| Mean CTD | Unspecified | Unspecified | Specified | Unspecified | |
| Maximum CTD | Specified | Specified | Unspecified | Unspecified | |
| CLR[1,5] | Specified | | | Unspecified | Specified[2] |
| Feedback | Unspecified | | | Unspecified | Specified |

Notes:

1. For all service categories, the cell loss ratio may be unspecified for CLP=1.

2. Minimized for sources that adjust cell flow in response to control information.

3. May not be subject to connection admission control and user parameter control procedures.

4. Represents the maximum rate at which the source can send as controlled by the control information.

5. These parameters are either explicitly or implicitly specified.

6. Different values of CDVT may specified for SCR and PCR.

# 3 Service Categories

There are five categories of service. The QoS parameters for these categories are summarized in Table 1 and are explained below [46]:

1. **Constant Bit Rate (CBR)**: This category is used for emulating circuit switching. The cell rate is constant. Cell loss ratio is specified for CLP=0 cells and may or may not be specified for CLP=1 cells. Examples of applications that can use CBR are telephone, video conferencing, and television (entertainment video).

2. **Variable Bit Rate (VBR)**: This category allows users to send at a variable rate. Statistical multiplexing is used and so there may be a small nonzero random loss. Depending upon whether or not the application is sensitive to cell delay variation, this category is subdivided into two categories: Real time VBR and Nonreal time VBR. For nonreal time VBR, only mean delay is specified, while for realtime VBR, maximum delay and peak-to-peak CDV are specified. An example of realtime VBR is interactive compressed video while that of nonreal time VBR is multimedia email.

3. **Available Bit Rate (ABR)**: This category is designed for normal data traffic such as file transfer and email. Although, the standard does not require the cell transfer delay and cell loss ratio to be guaranteed or minimized, it is desirable for switches to minimize the delay and loss as much as possible. Depending upon the congestion state of the network, the source is required to control its rate. The users are allowed to declare a minimum cell rate, which is guaranteed to the VC by the network. Most VCs will ask for an MCR of zero. Those with higher MCR may be denied connection if sufficient bandwidth is not available.

4. **Unspecified Bit Rate (UBR)**: This category is designed for those data applications that want to use any left-over capacity and are not sensitive to cell loss or delay. Such connections are not rejected on the basis of bandwidth shortage (no connection admission control) and not policed for their usage behavior. During congestion, the cells are lost but the sources are not expected to reduce their cell rate. In stead, these applications may have their own higher-level cell loss recovery and retransmission mechanisms. Examples of applications that can use this service are email, file transfer, news feed, etc. Of course, these same applications can use the ABR service, if desired.

Note that only ABR traffic responds to congestion feedback from the network. The rest of this paper is devoted to this category of traffic.

# 4  Congestion Control Methods

Congestion happens whenever the input rate is more than the available link capacity:

$$\sum \text{Input Rate} > \text{Available link capacity}$$

Most congestion control schemes consist of adjusting the input rates to match the available link capacity (or rate). One way to classify congestion control schemes is by the layer of ISO/OSI reference model at which the scheme operates. For example, there are data link, routing, and transport layer congestion control schemes. Typically, a combination of such schemes is used. The selection depends upon the severity and duration of congestion.

Figure 1 shows how the duration of congestion affects the choice of the method. The best method for networks that are almost always congested is to install higher speed links and redesign the topology to match the demand pattern.

For sporadic congestion, one method is to route according to load level of links and to reject new connections if all paths are highly loaded. This is called "connection admission control (CAC)." The "busy" tone on telephone networks is an example of CAC. CAC is effective only for medium duration congestion since once the connection is admitted the congestion may persist for the duration of the connection.

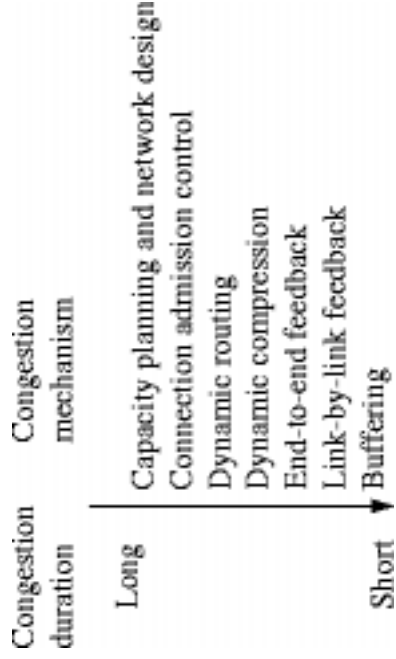

Figure 1: Congestion techniques for various congestion durations

For congestions lasting less than the duration of connection, an end-to-end control scheme can be used. For example, during connection setup, the sustained and peak rate may be negotiated. Later a leaky bucket algorithm may be used by the source or the network to ensure that the input meets the negotiated parameters. Such "traffic shaping algorithms" are open loop in the sense that the parameters cannot be changed dynamically if congestion is detected after negotiation. In a closed loop scheme, on the other hand, sources are informed dynamically about the congestion state of the network and are asked to increase or decrease their input rate. The feedback may be used hop-by-hop (at datalink layer) or end-to-end (transport layer). Hop-by-hop feedback is more effective for shorter term overloads than the end-to-end feedback.

For very short spikes in traffic load, providing sufficient buffers in the switches is the best solution.

Notice that solutions that are good for short term congestion are not good for long-term overload and vice-versa. A combination of various techniques (rather than just one technique) is used since overloads of various durations are experienced on all networks.

UNI 3.0 allows CAC, traffic shaping, and binary feedback (EFCI). However, the algorithms for CAC are not specified. The traffic shaping and feedback mechanisms are described next.

## 4.1  Generalized Cell Rate Algorithm (GCRA)

As discussed earlier, GCRA is the so called "leaky bucket" algorithm, which is used to enforce regularity in the cell arrival times. Basically, all arriving cells are put into a bucket, which is drained at the specified rate. If too many cells arrive at once, the bucket may overflow. The overflowing cells are called non-conforming and may or may not be admitted in to the network. If admitted, the cell loss priority (CLP) bit of the non-conforming cells may be set so that they will be first to be dropped in case of overload. Cells of service categories specifying peak cell rate should conform to GRCA(1/PCR, CDVT), while those also specifing sustained cell rate should additionally conform to GCRA(1/SCR, BT). See

Section 2 for definitions of CDVT and BT.

## 4.2 Feedback Facilities

UNI V3.0 specified two different facilities for feedback control:

1. Generalized Flow Control (GFC)
2. Explicit Forward Congestion Indication (EFCI)

| Generalized Flow Control | Virtual Path ID | Virtual Circuit ID | Payload Type | Cell Loss Priority | Header Error Check |
|---|---|---|---|---|---|
| 4 | 8 | 16 | 3 | 1 | 8 ← Size in bits |

Figure 2: ATM Cell header format

As shown in Figure 2, the first four bits of the cell header at the user-network interface (UNI) were reserved for GFC. At network-network interface, GFC is not used and the four bits are part of an extended virtual path (VP) field. It was expected that the GFC bits will be used by the network to flow control the source. The GFC algorithm was to be designed later. This approach has been abandoned.

Switches can use the payload type field in the cell header to convey congestion information in a binary (congestion or no congestion) manner. When the first bit of this field is 0, the second bit is treated as "explicit forward congestion indication (EFCI) bit." For all cells leaving their sources, the EFCI bit is clear. Congested switches on the path, if any, can set the EFCI bit. The destinations can monitor EFCI bits and ask sources to increase or decrease their rate. The exact algorithm for use of EFCI was left for future definition. The use of EFCI is discussed later in Section 8.

## 5 Selection Criteria

ATM network design started intially in CCITT (now known as ITU). However, the progress was rather slow and also a bit "voice-centric" in the sense that many of the decisions were not suitable for data traffic. So in October 1991, four companies – Adaptive (NET), CISCO, Northern Telecom, and Sprint, formed ATM Forum to expedite the process. Since then ATM Forum membership has grown to over 200 principal members. The traffic management working group was started in the Forum in May 1993. A number of congestion schemes were presented. To sort out these proposals, the group decided to first agree on a set of selection criteria. Since these criteria are of general interest and apply to non-ATM networks as well, we describe some of them briefly here.

## 5.1 Scalability

Networks are generally classified based on extent (coverage), number of nodes, speed, or number of users. Since ATM networks are intended to cover a wide range along all these dimensions, it is necessary that the scheme be not limited to a particular range of speed, distance, number of switches, or number of VCs. In particular, this ensures that the same scheme can be used for local area networks (LANs) as well as wide area networks (WANs).

## 5.2 Optimality

In a shared environment the throughput for a source depends upon the demands by other sources. The most commonly used criterion for what is the correct share of bandwidth for a source in a network environment, is the so called "max-min allocation [12]." It provides the maximum possible bandwidth to the source receiving the least among all contending sources. Mathematically, it is defined as follows. Given a configuration with n contending sources, suppose the $i$th source gets a bandwidth $x_i$. The allocation vector $\{x_1, x_2, \ldots, x_n\}$ is feasible if all link load levels are less than or equal to 100%. The total number of feasible vectors is infinite. For each allocation vector, the source that is getting the least allocation is in some sense, the "unhappiest source." Given the set of all feasible vectors, find the vector that gives the maximum allocation to this unhappiest source. Actually, the number of such vectors is also infinite although we have narrowed down the search region considerably. Now we take this "unhappiest source" out and reduce the problem to that of remaining n-1 sources operating on a network with reduced link capacities. Again, we find the unhappiest source among these n-1 sources, give that source the maximum allocation and reduce the problem by one source. We keep repeating this process until all sources have been given the maximum that they could get.

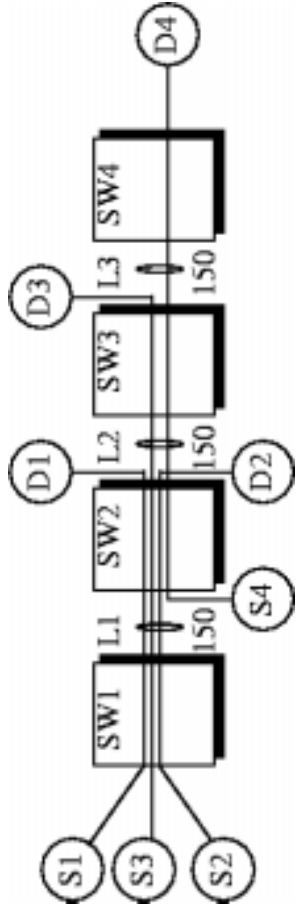

Figure 3: Sample configuration for max-min fairness

The following example illustrates the above concept of max-min fairness. Figure 3 shows a network with four switches connected via three 150 Mbps links. Four VCs are setup such that the first link L1 is shared by sources S1, S2, and S3. The second link is shared by S3 and S4. The third link is used only by S4. Let us divide the link bandwidths fairly among contending sources. On link L1, we can give 50 Mbps to each of the three contending sources S1, S2, and S3. On link L2, we would give 75 Mbps to each of the sources S3 and S4. On link L3, we would give all 155 Mbps to source S4. However, source S3 cannot use its 75 Mbps share at link L2 since it is allowed to use only 50 Mbps at link L1. Therefore, we give 50 Mbps to source S3 and construct a new configuration shown in Figure 4, where Source S3 has been removed and the link capacities have been reduced accordingly. Now we give 1/2 of the link L1's remaining capacity to each of the two contending sources: S1 and S2; each gets 50 Mbps. Source S4 gets the entire remaining bandwidth (100 Mbps) of link L2. Thus, the fair allocation vector for this configuration is (50, 50, 50, 100). This is the max-min allocation.

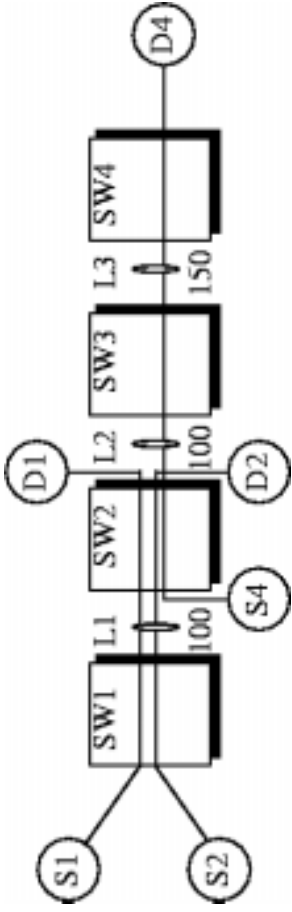

Figure 4: Configuration after removing VC 3.

Notice that max-min allocation is both fair and efficient. It is fair in the sense that all sources get an equal share on every link provided that they can use it. It is efficient in the sense that each link is utilized to the maximum load possible.

It must be pointed out that the max-min fairness is just one of several possible optimality criteria. It does not account for the guaranteed minimum (MCR). Other criterion such as weighted fairness have been proposed to determine optimal allocation of resources over and above MCR.

## 5.3 Fairness Index

Given any optimality criterion, one can determine the optimal allocation. If a scheme gives an allocation that is different from the optimal, its unfairness is quantified numerically as follows. Suppose a scheme allocates $\{\tilde{x}_1, \tilde{x}_2, ..., \tilde{x}_n\}$ instead of the optimal allocation $\{\hat{x}_1, \hat{x}_2, ..., \hat{x}_n\}$. Then, we calculate the normalized allocations $x_i = \tilde{x}_i/\hat{x}_i$ for each source and compute the fairness index as follows [13, 14]:

$$\text{Fairness} = \frac{(\sum_i x_i)^2}{n \sum_i x_i^2}$$

Since allocations $x_i$s usually vary with time, the fairness can be plotted as a function of time. Alternatively, throughputs over a given interval can be used to compute overall fairness.

## 5.4 Robustness

The scheme should be insensitive to minor deviations. For example, slight mistuning of parameters or loss of control messages should not bring the network down. It should be possible to isolate misbehaving users and protect other users from them.

## 5.5 Implementability

The scheme should not dictate a particular switch architecture. As discussed later in Section 9, this turned out to be an important point in final selection since many schemes were found to not work with FIFO scheduling.

## 5.6 Simulation Configurations

A number of network configuration were also agreed upon to compare various proposals. Most of these were straight forward serial connection of switches. The most popular one is the so called "Parking Lot" configuration for studying fairness. The configuration and its name is derived from theatre parking lots, which consist of several parking areas connected via a single exit path as shown in Figure 5. At the end of the show, congestion occurs as cars exiting from each parking area try to join the main exit stream.

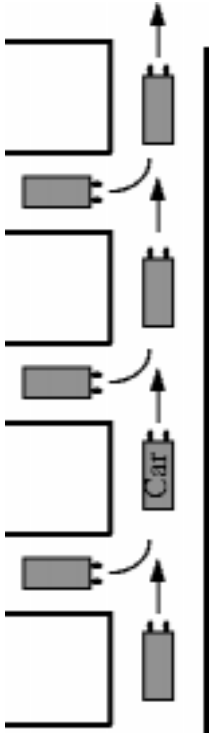

Figure 5: Theatre parking lot

For computer networks, an $n$-stage parking lot configuration consists of $n$ switches connected in a series. There are $n$ VCs. The first VC starts from the first switch and goes to the end. For the remaining $i$th VC starts at the $i - 1$th switch. A 3-switch parking lot configuration is shown in Figure 6.

## 5.7 Traffic Patterns

Among the traffic patterns used in various simulations, the following three were most common:

1. **Persistent Sources**: These sources, also known as "greedy" or "infinite" sources always have cells to send. Thus, the network is always congested.

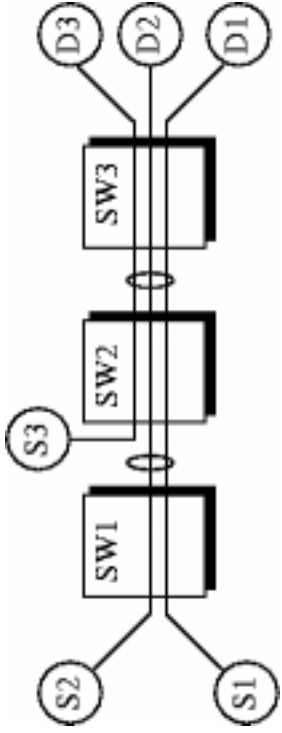

Figure 6: Parking lot configuration

2. **Staggered Source**: The sources start at different times. This allows us to study the ramp-up (or ramp-down) time of the schemes.

3. **Bursty Sources**: These sources oscillate between active state and idle state. During active state, they generate a burst of cells [10]. This is a more realistic source model than a persistent source. With bursty sources, if the total load on the link is less than 100%, then throughput and fairness are not at issue, what is more important is the "burst response time" – the time from "first cell in" to "the last cell out." If the bursts arrive at a fixed rate, it is called "open loop." A more realistic scenario is when the next burst arrives some time after the response to the previous burst has been received. In this later case, the burst arrival is affected by network congestion and so the traffic model is called "closed loop."

# 6 Congestion Schemes

In this section, we briefly describe proposals that were presented but were discarded early at the ATM Forum. The two key proposals – the credit based and the rate based – that were discussed at length are described in detail in the next two sections.

## 6.1 Fast Resource Management

This proposal from France Telecom [4] requires sources to send a resource management (RM) cell requesting the desired bandwidth before actually sending the cells. If a switch cannot grant the request it simply drops the RM cell; the source times out and resends the request. If a switch can satisfy the request, it passes the RM cell on to the next switch. Finally, the destination returns the cell back to the source which can then transmit the burst.

As described above, the burst has to wait for at least one round trip delay at the source even if the network is idle (as is often the case). To avoid this delay, an "immediate transmission (IT)" mode was also proposed in which the burst is transmitted immediately following the RM cell. If a switch cannot satisfy the request, it drops the cell and the burst and sends an indication to the source.

If cell loss, rather than bandwidth is of concern, the resource request could contain the burst size. A switch would accept the request only if it had that many buffers available.

The fast resource management proposal was not accepted at the ATM Forum primarily because it would either cause excessive delay during normal operation or excessive loss during congestion.

## 6.2 Delay-Based Rate Control

This proposal made by Fujitsu [24] requires that the sources monitor the round trip delay by periodically sending resource management (RM) cells that contain timestamp. The cells are returned by the destination. The source uses the timestamp to measure the roundtrip delay and to deduce the level of congestion. This approach, which is similar to that described in Jain [17], has the advantage that no explicit feedback is expected from the network and, therefore, it will work even if the path contained non-ATM networks.

Although the proposal was presented at the ATM Forum, it was not followed up and the precise details of how the delay will be used were not presented. Also, this method does not really require any standardization, since any source-destination pair can do this without involving the network.

## 6.3 Backward Explicit Congestion Notification (BECN)

This method presented by N.E.T. [28, 29, 30] consists of switches monitoring their queue length and sending an RM cell back to source if congested. The sources reduce their rates by half on the receipt of the RM cell. If no BECN cells are received within a recovery period, the rate for that VC is doubled once each period until it reaches the peak rate. To achieve fairness, the source recovery period was made proportional to the VC's rate so that lower the transmission rate the shorter the source recovery period.

This scheme was dropped because it was found to be unfair. The sources receiving BECNs were not always the ones causing the congestion [32].

## 6.4 Early Packet Discard

This method presented by Sun Microsystems [35] is based on the observation that a packet consists of several cells. It is better to drop all cells of one packet then to randomly drop cells belonging to different packets. In AAL5, when the first bit of the payload type bit in the cell header is 0, the third bit indicates "end of message (EOM)." When a switch's queues start getting full, it looks for the EOM marker and it drops all future cells of the VC until the "end of message" marker is seen again.

It was pointed out [33] that the method may not be fair in the sense that the cell to arrive at a full buffer may not belong to the VC causing the congestion.

Note that this method does not require any inter-switch or source-switch communication and, therefore, it can be used without any standardization. Many switch vendors are implementing it.

## 6.5 Link Window with End-to-End Binary Rate

This method presented by Tzeng and Siu [45], consisted of combining good features of the credit-based and rate-based proposals being discussed at the time. It consists of using window flow control on every link and to use binary (EFCI-based) end-to-end rate control. The window control is per-link (and not per-VC as in credit-based scheme). It is, therefore, scalable in terms of number of VCs and guarantees zero cell loss. Unfortunately, neither the credit-based nor the rate-based camp found it acceptable since it contained elements from the opposite camp.

## 6.6 Fair queueing with Rate and Buffer feedback

This proposal from Xerox and CISCO [27] consists of sources periodically sending RM cells to determine the bandwidth and buffer usage at their bottlenecks. The switches compute fair share of VCs. The minimum of the share at this switch and that from previous switches is placed in the RM cells. The switches also monitor each VC's queue length. The maximum of queue length at this switch and those from the previous switches is placed in the same RM cell. Each switch implements fair queueing, which consists of maintaining a separate queue for each VC and computing the time at which the cell would finish transmission if the queues were to be served round-robin one-bit at a time. The cells are scheduled to transmit in this computed time order.

The fair share of a VC is determined as the inverse of the interval between the cell arrival and its transmission. The interval reflects the number of other VCs that are active. Since the number and hence the interval is random, it was recommended that the average of several observed interval be used.

This scheme requires per-VC (fair) queueing in the switches, which was considered rather complex.

# 7 Credit-Based Approach

This was one of the two leading approaches and also the first one to be proposed, analyzed, and implemented. Originally proposed by Professor H. T. Kung, it was supported by Digital, BNR, FORE, Ascom-Timeplex, SMC, Brooktree, and Mitsubishi [26, 41]. The approach consists of per-link, per-VC, window flow control. Each link consists of a sender node (which can be a source end system or a switch) and a receiver node (which can be a switch or a destination end system). Each node maintains a separate queue for each VC. The receiver monitors queue lengths of each VC and determines the number of cells that the sender can transmit on that VC. This number is called "credit." The sender transmits only as many cells as allowed by the credit.

If there is only one active VC, the credit must be large enough to allow the whole link to be full at all times. In other words:

$$\text{Credit} \geq \text{Link Cell Rate} \times \text{Link Round Trip Propagation Delay}$$

The link cell rate can be computed by dividing the link bandwidth in Mbps by the cell size in bits.

The scheme as described so far is called "Flow Controlled Virtual Circuit (FCVC)" scheme. There are two problems with this initial static version. First, if the credits are lost, the sender will not know it. Second each VC needs to reserve the entire round trip worth of buffers even though the link is shared by many VCs. These problems were solved by introducing a credit resynchronization algorithm and an adaptive version of the scheme.

The credit resynchronization algorithm consists of both sender and receiver maintaining counts of cells sent and received for each VC and periodically exchanging these counts. The difference between the cells sent by the sender and those received by the receiver represents the number of cells lost on the link. The receiver reissues that many additional credits for that VC.

The adaptive FCVC algorithm [25] consists of giving each VC only a fraction of the roundtrip delay worth of buffer allocation. The fraction depends upon the rate at which the VC uses the credit. For highly active VCs, the fraction is larger while for less active VCs, the fraction is smaller. Inactive VCs get a small fixed credit. If a VC doesn't use its credits, its observed usage rate over a period is low and it gets smaller buffer allocation (and hence credits) in the next cycle. The adaptive FCVC reduces the buffer requirements considerably but also introduces a ramp-up time. If a VC becomes active, it may take some time before it can use the full capacity of the link even if there are no other users.

# 8 Rate-Based Approach

This approach, which was eventually adopted as the standard was proposed originally by Mike Hluchyj and was extensively modified later by representatives from 22 different companies [7].

Original proposal consisted of a rate-based version of the DECbit scheme [18], which consists of end-to-end control using a single-bit feedback from the network. In the proposal, the switches monitor their queue lengths and if congested set EFCI in the cell headers. The destination monitors these indications for a periodic interval and sends an RM cell back to the source. The sources use an additive increase and multiplicative decrease algorithm to adjust their rates.

This particular algorithm uses a "negative polarity of feedback" in the sense that RM cells are sent only to decrease the rate but no RM cells are required to increase the rate. A positive polarity, on the other hand, would require sending RM cells for increase but not on decrease. If RM cells are sent for both increase and decrease, the algorithm would be called bipolar.

The problem with negative polarity is that if the RM cells are lost due to heavy congestion in the reverse path, the sources will keep increasing their load on the forward path and eventually overload it.

This problem was fixed in the next version by using positive polarity. The sources set EFCI on every cell except the nth cell. The destination will send an "increase" RM cell to source if they receive any cells with the EFCI off. The sources keep decreasing their rate until they receive a positive feedback. Since the sources decrease their rate proportional to the current rate, this scheme was called "proportional rate control algorithm (PRCA)."

PRCA was found to have a fairness problem. Given the same level of congestion at all switches, the VCs travelling more hops have a higher probability of having EFCI set than those travelling smaller number of hops. If p is the probability of EFCI being set on one hop, then the probability of it being set for an n-hop VC is $1 - (1-p)^n$ or $np$. Thus, long path VCs have fewer opportunities to increase and are beaten down more often than short path VCs. This was called the "beat-down problem [3]."

One solution to the beat down problem is the selective feedback [31] or intelligent marking [1] in which a congested switch takes into account the current rate of the VC in addition to its congestion level in deciding whether to set the EFCI in the cell. The switch computes a "fair share" and if congested it sets EFCI bits in cells belonging to only those VCs whose rates are above this fair share. The VCs with rates below fair share are not affected.

## 8.1 The MIT Scheme

In July 1994, we [6] argued that the binary feedback was too slow for rate-based control in high-speed networks and that an explicit rate indication would not only be faster but would offer more flexibility to switch designers.

The single-bit binary feedback can only tell the source whether it should go up or down. It was designed in 1986 for connectionless networks in which the intermediate nodes had no knowledge of flows or their demands. The ATM networks are connection oriented. The switches know exactly who is using the resources and the flow paths are rather static. This increased information is not used by the binary feedback scheme.

Secondly and more importantly, the binary feedback schemes were designed for window-based controls and are too slow for rate-based contols. With window-based control a slight difference between the current window and the optimal window will show up as a slight increase in queue length. With rate-based control, on the other hand, a slight difference in current rate and the optimal rate will show up as continuously increasing queue length [15, 16]. The reaction times have to be fast. We can no longer afford to take several round trips that the binary feedback requires to settle to the optimal operation. The explicit rate feedback can get the source to the optimal operating point within a few round trips.

The explicit rate schemes have several additional advantages. First, policing is straight forward. The entry switches can monitor the returning RM cells and use the rate directly in their policing algorithm. Second with fast convergence time, the system come to the optimal operating point quickly. Initial rate has less impact. Third, the schemes are robust against errors in or loss of RM cells. The next correct RM cell will bring the system to the correct operating point.

We substantiated our arguments with simulation results for an explicit rate scheme designed by Anna Charny during her master thesis work at the Massachusetts Institute of Technology (MIT) [5]. The MIT scheme consists of each source sending an RM cell every nth data cell. The RM cell contains the VC's current cell rate (CCR) and a "desired rate." The switches monitor all VC's rates and compute a "fair share." Any VC's whose desired rate is less than the fair share is granted the desired rate. If a VC's desired rate is more than the fair share, the desired rate field is reduced to the fair share and a "reduced bit" is set in the RM cell. The destination returns the RM cell back to the source, which then adjusts its rate to that indicated in the RM cell. If the reduced bit is clear, the source could demand a higher desired rate in the next RM cell. If the bit is set, the source use the current rate as the desired rate in the next RM cell.

The fair share is computed using a iterative procedure as follows. Initially, the fair share is set at the link bandwidth divided by the number of active VC's. All VCs, whose rates are less than the fair share are called "underloading VCs". If the number of underloading VCs increases at any iteration, the fair share is recomputed as follows:

$$\text{Fair Share} = \frac{\text{Link Bandwidth} - \sum \text{Bandwidth of Underloading VCs}}{\text{Number of VCs} - \text{Number of Underloading VCs}}$$

The iteration is then repeated until the number of underloading VCs and the fair share does not change. Charny [5] has shown that two iterations are sufficient for this procedure to converge.

Charny also showed that the MIT scheme achieve max-min optimality in $4k$ round trips, where $k$ is the number of bottlenecks.

This proposal was well received except that the computation of fair share requires order $n$ operations, where $n$ is the number of VCs. Search for an O(1) scheme led to the EPRCA algorithm discussed next.

## 8.2 Enhanced PRCA (EPRCA)

The merger of PRCA with explicit rate scheme lead to the "Enhanced PRCA (EPRCA)" scheme at the end of July 1994 ATM Forum meeting [36, 38]. In EPRCA, the sources send data cells with EFCI set to 0. After every $n$ data cells, they send an RM cell.

The RM cells contain desired explicit rate (ER), current cell rate (CCR), and a congestion indication (CI) bit. The sources initialize the ER field to their peak cell rate (PCR) and set the CI bit to zero.

The switches compute a fair share and reduce the ER field in the returning RM cells to the fair share if necessary. Using exponential weighted averaging a mean allowed cell rate (MACR) is computed and the fair share is set at a fraction of this average:

$$\text{MACR} = (1 - \alpha)\text{MACR} + \alpha\text{CCR}$$

$$\text{Fair Share} = \text{SW\_DPF} \times \text{MACR}$$

Here, $\alpha$ is the exponential averaging factor and SW_DPF is a multiplier (called switch down pressure factor) set close to but below 1. The suggested values of $\alpha$ and SW_DPF are 1/16 and 7/8, respectively.

The destinations monitor the EFCI bits in data cells. If the last seen data cell had EFCI bit set, they mark the CI bit in the RM cell.

In addition to setting the explicit rate, the switches can also set the CI bit in the returning RM cells if their queue length is more than a certain threshold.

The sources decrease their rates continuously after every cell.

$$\text{ACR} = \text{ACR} \times \text{RDF}$$

Here, RDF is the reduction factor. When a source receives the returned RM cell, it increases its rate by an amount AIR if permitted.

$$\text{IF CI=0 Then New ACR} = \text{Min(ACR+AIR, ER, PCR)}$$

If CI bit is set, the ACR is not changed.

Notice that EPRCA allows both binary-feedback switches and the explicit feedback switches on the path. The main problem with EPRCA as described here is the switch congestion detection algorithm. It is based on queue length threshold. If the queue length exceeds a certain threshold, the switch is said to be congested. If it exceed another higher threshold, it said to be very highly congested. This method of congestion detection was shown to result in unfairness. Sources that start up late were found to get lower throughput than those which start early.

The problem was fixed by changing to queue growth rate as the load indicator. The change in the queue length is noted down after processing, say, K cells. The overload is indicated if the queue length increases [44, 43].

## 8.3 OSU Time-based Congestion Avoidance

Jain, Kalyanaraman, and Viswanathan at the Ohio State University (OSU) have developed a series of explicit rate congestion avoidance schemes. The first scheme [19, 20] called the OSU scheme consists of switches measuring their input rate over a fixed "averaging interval" and comparing it with their target rate to compute the current Load factor $z$:

$$\text{Load Factor } z = \frac{\text{Input rate}}{\text{Target Rate}}$$

The target rate is set at slightly below, say, 85-95% of the link bandwidth. Unless the load factor is close to 1, all VCs are asked to change (divide) their load by this factor $z$. For example, if the load factor is 0.5, all VCs are asked to divide their rate by a factor of 0.5, that is, double their rates. On the other hand, if the load factor is 2, all VCs would be asked to halve their rates.

Note that no selective feedback is taken when the switch is either highly overloaded or highly underloaded. However, if the load factor is close to one, between 1-$\Delta$ and 1+$\Delta$ for a small $\Delta$, the switch gives different feedback to underloading sources and overloading sources. A fair share is computed as follows:

$$\text{Fair Share} = \frac{\text{Target Rate}}{\text{Number of Active Sources}}$$

All sources, whose rate is more than the fair share are asked to divide their rates by $z/(1+\Delta)$ while those below the fair share are asked to divide their rates by $z/(1-\Delta)$. This algorithm called "Target Utilization Band (TUB) algorithm" was proven to lead to fairness [19].

The OSU scheme has three distinguishing features. First, it is a congestion avoidance scheme. It gives high throughput and low delay. By keeping the target rate slightly below the capacity, the algorithm ensures that the queues are very small, typically close to 1, resulting in low delay. Second, the switches have very few parameters compared to EPRCA and are easy to set. Third, the time to reach the steady state is very small. The source reach their final operating point 10 to 20 times faster than that with EPRCA.

In the original OSU scheme, the sources were required to send RM cells periodically at fixed time interval. This meant that the RM cell overhead per source was fixed and increased as the number of sources increases. This was found to be unacceptable leading to the count-based scheme described next.

In the count-based scheme [21], the sources send RM cells after every $n$ data cells, as in EPRCA. The switch rate adjustment algorithm is changed to encourage quick rise. Sources below the fair share are asked to come up to the fair share regardless of the load level and those above the fair share are asked to adjust their rates by the load factor. This allows the scheme to keep the three distinguishing feature while making the overhead independent of number of VCs. Newer versions of the OSU scheme, named "ERICA" (Explicit Congestion Indication for Congestion Avoidance) and "ERICA+" are count-based [22, 23].

## 8.4 Congestion Avoidance using Proportional Control (CAPC)

Andy Barnhart from Hughes Systems has proposed a scheme called "Congestion Avoidance using Proportional Control (CAPC)[2]." In this scheme, as in OSU scheme, the switches set a target utilization slightly below 1. This helps keep the queue length small. The switches measure the input rate and load factor $z$, as in OSU scheme, and use it to update the fair share.

During underload ($z < 1$), fair share is increased as follows:

$$\text{Fair share} = \text{Fair share} \times \text{Min}(ERU, 1 + (1 - z) * Rup)$$

Here, Rup is a slope parameter in the range 0.025 to 0.1. ERU is the maximum increase allowed and was set to 1.5.

During overload ($z > 1$), fair share is decreased as follows:

$$\text{Fair share} = \text{Fair share} \times \text{Max}(ERF, 1 - (z - 1) * Rdn)$$

Here, Rdn is a slope parameter in the range 0.2 to 0.8 and ERF is the minimum decrease required and was set to 0.5.

The fair share is the maximum rate that the switch will grant to any VC.

In addition to the load factor, the scheme also uses a queue threshold. Whenever the queue length is over this threshold, a congestion indication (CI) bit is set in all RM cells. This prevents all sources from increasing their rate and allows the queues to drain out.

The distinguishing feature of CAPC is oscillation-free steady state performance. The frequency of oscillations is a function of $1 - z$, where $z$ is the load factor. In steady state, $z = 1$, the frequency is zero, that is, the period of oscillations is infinite. This scheme is still under development.

It must be pointed out that the final specification does not mandate any particular switch algorithm. EPRCA, ERICA, CAPC, and a few other algorithms are included in the appendix as examples of possible switch algorithms. However, each manufacturer is free to use its own algorithm.

## 8.5 Virtual Source and Destination

One objection to the end-to-end rate control is that the round trip delay can be very large. This problem is fixed by segmenting the network in smaller pieces and letting the switches act as "virtual source" and/or "virtual destination." Figure 7 shows an example [8]. Switch A in the middle segment acts as a virtual destination and returns all RM cells received from the source as if the switch was the destination. Switch B in the same segment acts as a virtual source and generates RM cells as if it were a source.

Segmenting the network using virtual source/destination reduces the size of the feedback loops. Also, the intermediate segments can use any proprietary congestion control scheme. This allows public telecommunications carriers to follow the standard interface only at entry/exit switches. More importantly, virtual source/destination provide a more robust interface to a public network in the sense that the resources inside the network do not have to rely on user compliance. Misbehaving users will be isolated in the first control loop. The users here include private networks with switches that may or may not be compliant.

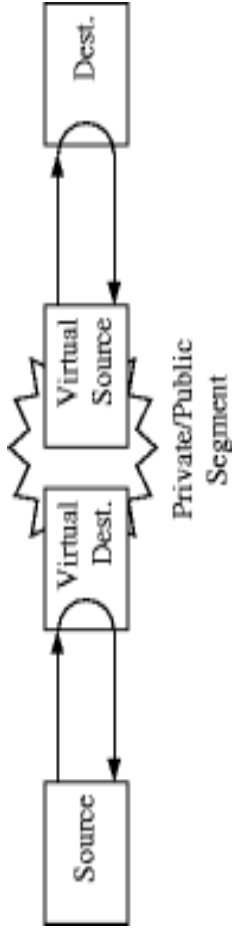

Figure 7: Virtual source/destination.

Notice that the virtual sources and destinations need to maintain per-VC queueing and may, therefore, be quite expensive.

There is no limit on the number of segments that can be created. In the extreme case, every switch could act as a virtual source/destination and one would get "hop-by-hop" rate control as shown in Figure 8.

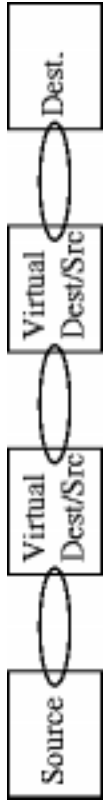

Figure 8: Hop-by-hop rate control.

### 8.6 Multicast VCs

The explicit rate approach can be extended for point-to-multipoint connections. As shown in Figure 9, the forward RM cells are copied to each branch. While in the reverse direction, the returning RM cell information is combined so that the minimum of the rates allowed by the branches is fed back towards the root. In one such schemes, proposed by Roberts [39], the information from the returning RM cells is kept at the branching node and is returned whenever the next forward RM cell is received.

## 9 Credit vs Rate Debate

After a considerable debate [9, 11, 34, 37], which lasted for over a year, ATM Forum adopted the rate-based approach and rejected the credit-based approach. The debate was quite "religious" in the sense that believers of each approach had quite different goals in mind and were unwilling to compromise. To achieve their goals, they were willing to make tradeoffs that were unacceptable to the other side. In this section, we summarize some of the key points that were raised during this debate.

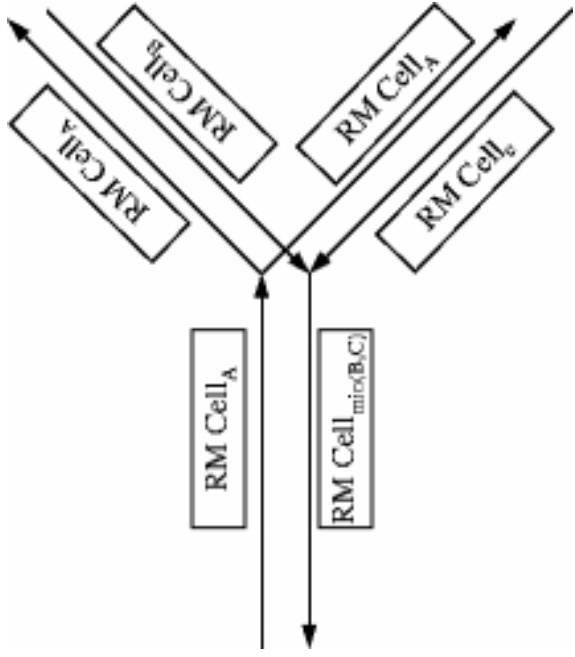

Figure 9: Explicit rate control for multicast VCs

1. **Per-VC Queueing**: Credit-based approach requires switches to keep a separate queue for each VC (or VP in case of virtual path scheduling). This applies even to inactive VCs. Per-VC queueing makes switch complexity proportional to the number of VCs. The approach was considered not scalable to a large number of VCs. Given that some large switches will support millions of VCs, this would cause considerable complexity in the switches. This was the single biggest objection to the credit-based approach and the main reason for it not being adopted. Rate-based approach does not require per-VC queueing. It can work with or without per-VC queueing. The choice is left to the implementers.

2. **Zero Cell Loss**: Under ideal conditions, the credit-based approach can guarantee zero cell loss regardless of traffic patterns, the number of connections, buffer sizes, number of nodes, range of link bandwidths, and range of propagation and queueing delays. Even under extreme overloads, the queue lengths cannot grow beyond the credits granted. The rate-based approach cannot guarantee cell loss. Under extreme overloads, it is possible for queues to grow large resulting in buffer overflow and cell loss. The rate-based camp considered the loss acceptable arguing that with large buffers, the probability of loss is small. Also, they argued that in reality there is always some loss due to errors and, therefore, the user has to worry about loss even if there is zero congestion loss.

3. **Ramp-Up Time**: The static credit-based approach allows VCs to ramp up to the full rate very fast. In fact, any free capacity can be used immediately. Some rate-based schemes (and the adaptive credit-based approach) can take several round trip delays to ramp up.

4. **Isolation and Misbehaving Users**: A side benefit of the per-VC queueing is that misbehaving users cannot disrupt the operation of well-behaving users. However, this is less true for the adaptive scheme than for the static credit scheme. In the adaptive scheme, a misbehaving user can get a higher share of buffers by increasing its rate. Note that isolation is attained by per-VC queueing and not so much by credits. Thus, if required, a rate-based switch can also achieve isolation by implementing per-VC queueing.

5. **Buffer Requirements**: The buffer requirements for the credit-based schemes were found to be less than those in the rate-based scheme with binary feedback. However, this advantage disappeared when explicit rate schemes were added. In the static credit-based approach, per-VC buffer requirement is proportional to link delay, while in the rate-based approach, total buffer requirement is proportional to the end-to-end delay. The adaptive credit scheme allows lower buffering requirements than the static scheme but at the cost of increased ramp-up time.

   Note that the queueing delays have to be added in both cases since it delays the feedback and adds to the reaction time.

6. **Delay estimate**: Setting the congestion control parameters in the credit-based approach requires knowledge of link round trip delay. At least, the link length and speed must be known. This knowledge is not required for rate-based approaches (although it may be helpful).

7. **Switch Design Flexibility**: The explicit rate schemes provide considerable flexibility to switches in deciding how to allocate their resources. Different switches can use different mechanisms and still interoperate in the same network. For example, some switches can opt for minimizing their queue length, while the others can optimize their throughput, while still others can optimize their profits. On the other hand, the credit-based approach dictated that each switch use per-VC queueing with round-robin service.

8. **Switch vs End-System Complexity**: The credit-based approach introduces complexity in the switches but may have made the end-system's job a bit simpler. The proponents of credit-based approach argued that their host network interface card (NIC) is much simpler since they do not have to schedule each and every cell. As long as credits are available, the cells can be sent at the peak rate. The proponents of the rate-based approach countered that all NIC cards have to have schedulers for their CBR and VBR traffic and using the same mechanism for ABR does not introduce too much complexity.

There were several other points that were raised. But all of them are minor compared to the per-VC queueing required by the credit-based approach. Majority of the ATM Forum participants were unwilling to accept any per-VC action as a requirement in the switch. This is why, even an integrated proposal allowing vendors to choose either of the two approaches failed. The rate-based approach won by a vote of 7 to 104.

## 10 Summary

Congestion control is important in high speed networks. Due to larger bandwidth-distance product, the amount of data lost due to simultaneous arrivals of bursts from multiple sources can be larger. For the success of ATM, it is important that it provides a good traffic management for both bursty and non-bursty sources.

Based on the type of the traffic and the quality of service desired, ATM applications can use one of the five service categories: CBR, rt-VBR, nrt-VBR, UBR, and ABR. Of these, ABR is expected to be the most commonly used service category. It allows ATM networks to control the rates at which delay-insensitive data sources may transmit. Thus, the link bandwidth not used by CBR and VBR applications can be fairly divided among ABR sources.

After one year of intense debate on ABR traffic management, two leading approaches emerged: credit-based and rate-based. The credit-based approach uses per-hop per-VC window flow control. The rate-based approach uses end-to-end rate-based control. The rate-based approach was eventually selected primarily because it doesn't require all switches to keep per-VC queueing and allows considerable flexibility in resource allocation.

Although, the ATM Forum traffic management version 4.0 specifications allows older EFCI switches with binary feedback for backward compatibility, the newer explicit rate switches will provide better performance and faster control. The ATM Forum has specified a number of rules that end-systems and switches follow to adjust the ABR traffic rate. The specification is expected to be finalized by February 1996 with products appearing in mid to late 1996.

## 11 Acknowledgement

Thanks to Dr. Shirish Sathaye of FORE Systems, Dr. David Hughes of StrataCom, and Rohit Goyal and Ram Viswanathan of OSU for useful feedback on an earlier version of this paper.

## References

[1] A. W. Barnhart, "Use of the Extended PRCA with Various Switch Mechanisms," AF-TM [1] 94-0898, September 1994.

[2] A. W. Barnhart, "Explicit Rate Performance Evaluation," AF-TM 94-0983R1, October 1994.

---

[1]Throughout this section, AF-TM refers to ATM Forum Traffic Management working group contributions. These contributions are available only to Forum member organizations. All our contributions and papers are available on-line at http://www.cis.ohio-state.edu/~jain/

[3] J. Bennett and G. Tom Des Jardins, "Comments on the July PRCA Rate Control Baseline," AF-TM 94-0682, July 1994.

[4] P. E. Boyer and D. P. Tranchier, "A Reservation Principle with Applications to the ATM Traffic Control," Computer Networks and ISDN Systems, Vol. 24, 1992, pp. 321-334.

[5] A. Charny, "An Algorithm for Rate Allocation in a Cell-Switching Network with Feedback", MIT TR-601, May 1994.

[6] A. Charny, D. Clark, and R. Jain, "Congestion Control with Explicit Rate Indication," AF-TM 94-0692, July 1994. Updated version published in Proc. ICC'95, June 1995.

[7] M. Hluchyj, et al., "Closed-Loop Rate-Based Traffic Management," AF-TM 94-0211R3, April 1994.

[8] M. Hluchyj et al, "Closed-Loop Rate-Based Traffic Management," AF-TM 94-0438R2, September 1994.

[9] D. Hughes and P. Daley, "Limitations of Credit Based Flow Control," AF-TM 94-0776, September 1994.

[10] D. Hughes and P. Daley, "More ABR Simulation Results," AF-TM 94-0777, September 1994.

[11] D. Hunt, Shirish Sathaye, and K. Brinkerhoff, "The Realities of Flow Control for ABR Service," AF-TM 94-0871, September 1994.

[12] J. Jaffe, "Bottleneck Flow Control," IEEE Trans. Comm., Vol. COM-29, No. 7, pp. 954-962.

[13] R. Jain, D. Chiu, and W. Hawe, "A Quantitative Measure of Fairness and Discrimination for Resource Allocation in Shared Systems," DEC TR-301, September 1984.

[14] R. Jain, "The Art of Computer Systems Performance Analysis," John Wiley & Sons, New York, 1991.

[15] R. Jain, "Myths about Congestion Management in High Speed Networks," *Internetworking: Research and Experience*, Vol 3, 1992, pp. 101-113.

[16] R. Jain, "Congestion Control in Computer Networks: Issues and Trends," *IEEE Network Magazine*, May 1990, pp. 24-30.

[17] R. Jain, "A Delay-Based Approach for Congestion Avoidance in Interconnected Heterogeneous Computer Networks," *Computer Communications Review*, Vol. 19, No. 5, October 1989, pp. 56-71.

[18] R. Jain, K. K. Ramakrishnan, and D. M. Chiu, "Congestion Avoidance in Computer Networks with a Connectionless Network Layer," Digital Equipment Corporation, Technical Report, DEC-TR-506, August 1987, 17 pp. Also in C. Partridge, Ed., *Innovations in Internetworking*, Artech House, Norwood, MA, 1988, pp. 140-156.

[19] R. Jain, S. Kalyanaraman, and R. Viswanathan, "The OSU Scheme for Congestion Avoidance Using Explicit Rate Indication," AF-TM 94-0883, September 1994.

[20] R. Jain, S. Kalyanaraman, and R. Viswanathan, "The EPRCA+ Scheme" AF-TM 94-0988, October 1994.

[21] R. Jain, S. Kalyanaraman, and R. Viswanathan, "The Transient Performance: EPRCA vs EPRCA++," AF-TM 94-1173, November 1994.

[22] R. Jain, S. Kalyanaraman, and R. Viswanathan, "A Sample Switch Algorithm," AF-TM 95-0178R1, February 1995.

[23] R. Jain, S. Kalyanaraman, R. Goyal, S. Fahmy, F. Lu, "ERICA+: Extensions to the ERICA Switch Algorithm," AF-TM 95-1145R1, October 1995.

[24] D. Kataria, "Comments on Rate-Based Proposal," AF-TM 94-0384, May 1994.

[25] H. T. Kung, et al, "Adaptive Credit Allocation for Flow-Controlled VCs," AF-TM 94-0282, March 1994

[26] H. T. Kung, et al, "Flow Controlled Virtual Connections Proposal for ATM Traffic Management," AF-TM 94-0632R2, September 1994.

[27] B. Lyles and A. Lin, "Definition and Preliminary Simulation f a Rate-based Congestion Control Mechanism with Explicit Feedback of Bottleneck Rates," AF-TM 94-0708, July 1994.

[28] P. Newman and G. Marshall, "BECN Congestion Control," AF-TM 94-789R1, July 1993.

[29] P. Newman and G. Marshall, "Update on BECN Congestion Control," AF-TM 94-855R1, September 1993.

[30] P. Newman, "Traffic Management for ATM Local Area Networks," IEEE Communications Magazine, August 1994, pp. 44-50.

[31] K. K. Ramakrishnan, D. M. Chiu, and R. Jain, "Congestion Avoidance in Computer Networks with a Connectionless Network Layer. Part IV: A Selective Binary Feedback Scheme for General Topologies," Digital Equipment Corporation, Technical Report DEC-TR-510, August 1987, 41 pp.

[32] K. K. Ramakrishnan, "Issues with Backward Explicit Congestion Notification based Congestion Control," 93-870, September 1993.

[33] K. K. Ramakrishnan and J. Zavgren, "Preliminary Simulation Results of Hop-by-hop/VC Flow Control and Early Packet Discard," AF-TM 94-0231, March 1994.

[34] K. K. Ramakrishnan and P. Newman, "Credit Where Credit is Due," AF-TM 94-0916, September 1994.

[35] A. Romanov, "A Performance Enhancement for Packetized ABR and VBR+ Data," AF-TM 94-0295, March 1994.

[36] L. Roberts, "Enhanced PRCA (Proportional Rate-Control Algorithm)," AF-TM 94-0735R1, August 1994.

[37] L. Roberts, "The Benefits of Rate-Based Flow Control for ABR Service," AF-TM 94-0796, September 1994.

[38] L. Roberts et al., "New Pseudocode for Explicit Rate Plus EFCI Support," AF-TM 94-0974, October 1994.

[39] L. Roberts, "Rate-based Algorithm for Point to Multipoint ABR Service," AF-TM 94-0772R1, November 1994.

[40] S. Sathaye, "Draft ATM Forum Traffic Management Specification Version 4.0," AF-TM 95-0013, December 18, 1994.

[41] J. Scott, et al, "Link by Link, Per VC Credit Based Flow Control," AF-TM 94-0168, March 1994.

[42] W. Stallings, "ISDN and Broadband ISDN with Frame Relay and ATM," Prentice-Hall, 1995, 581 pp.

[43] H. Tzeng and K. Siu, "A Class of Proportional Rate Control Schemes and Simulation Results," AF-TM 94-0888, September 1994.

[44] K. Siu and H. Tzeng, "Adaptive Proportional Rate Control for ABR Service in ATM Networks," University of California, Irvine, Technical Report 94-07-01.

[45] H. Tzeng and K. Siu, "Enhanced Credit-Based Congestion Notification (ECCN) Flow Control for ATM Networks," AF-TM 94-0450, May 1994.

[46] L. Wojnaroski, "Baseline text for Traffic Management Sub-Working Group," AF-TM 94-0394R4, September 1994.